\documentclass[twocolumn, preprintnumbers,prl,aps,amssymb,showpacs,superscriptaddress]{revtex4}

\usepackage{graphicx}
\usepackage{dcolumn}
\usepackage{bm}
\usepackage{color}

\begin{document}

\newcommand{\ie}{{\it i.e.}}
\newcommand{\eg}{{\it e.g.}}
\newcommand{\etal}{{\it et al.}}


\title{Doping dependence of heat transport in the iron-arsenide superconductor Ba(Fe$_{1-x}$Co$_x$)$_2$As$_2$:
from isotropic to strongly $k$-dependent gap structure}


\author{M.~A.~Tanatar}
\affiliation{Ames Laboratory, Ames, Iowa 50011, USA}

\author{J.-Ph.~Reid}
\affiliation{D\'epartement de physique
\& RQMP, Universit\'e de Sherbrooke, Sherbrooke, Canada}

\author{H.~Shakeripour}
\affiliation{D\'epartement de physique
\& RQMP, Universit\'e de Sherbrooke, Sherbrooke, Canada}

\author{X.~G.~Luo}
\affiliation{D\'epartement de physique
\& RQMP, Universit\'e de Sherbrooke, Sherbrooke, Canada}

\author{N.~Doiron-Leyraud} \affiliation{D\'epartement de physique
\& RQMP, Universit\'e de Sherbrooke, Sherbrooke, Canada}

\author{N.~Ni}
\affiliation{Ames Laboratory, Ames, Iowa 50011, USA}
\affiliation{Department of Physics and Astronomy, Iowa State
University, Ames, Iowa 50011, USA }

\author{S.~L.~Bud'ko}
\affiliation{Ames Laboratory, Ames, Iowa 50011, USA}
\affiliation{Department of Physics and Astronomy, Iowa State
University, Ames, Iowa 50011, USA }

\author{P.~C.~Canfield}
\affiliation{Ames Laboratory, Ames, Iowa 50011, USA}
\affiliation{Department of Physics and Astronomy, Iowa State
University, Ames, Iowa 50011, USA }

\author{R.~Prozorov}
 \affiliation{Ames Laboratory, Ames,
Iowa 50011, USA} \affiliation{Department of Physics and Astronomy,
Iowa State University, Ames, Iowa 50011, USA }

\author{Louis Taillefer}
\altaffiliation{E-mail: louis.taillefer@physique.usherbrooke.ca }
\affiliation{D\'epartement de physique \& RQMP, Universit\'e de
Sherbrooke, Sherbrooke, Canada} \affiliation{Canadian Institute for
Advanced Research, Toronto, Ontario, Canada}

\date{\today}


\begin{abstract}

The temperature and magnetic field dependence of the in-plane thermal conductivity $\kappa$ of the iron-arsenide superconductor Ba(Fe$_{1-x}$Co$_x$)$_2$As$_2$ was measured down to $T \simeq 50$~mK and up to $H = 15$~T as a function of Co concentration $x$ in the range 0.048~$ \leq x \leq $~0.114.
In zero magnetic field, a negligible residual linear term in $\kappa/T$ as $T \to 0$ at all $x$ shows that there are no zero-energy quasiparticles and hence the superconducting gap has no nodes in the $ab$-plane anywhere in the phase diagram.
However, the field dependence of $\kappa$ reveals a systematic evolution of the superconducting gap with doping $x$, from large everywhere on the Fermi surface in the underdoped regime, as evidenced by a flat $\kappa (H)$ at $T \to 0$,
to strongly $k$-dependent in the overdoped regime, where a small magnetic field can induce a large residual linear term, indicative of a deep minimum in the gap magnitude somewhere on the Fermi surface.
This shows that the superconducting gap structure has a strongly $k$-dependent amplitude around the Fermi surface only outside the antiferromagnetic/orthorhombic phase.

\end{abstract}

\pacs{74.25.Fy, 74.20.Rp,74.70.Dd}

\maketitle


The pairing symmetry of the order parameter in iron-arsenide superconductors remains a subject of debate as different experimental probes point to different conclusions about the gap structure \cite{Hosono-review}. For example, while the power-law dependence of the penetration depth \cite{Martin2} and rapid increase of thermal conductivity with magnetic field \cite{BaKthermcond} in (Ba$_{1-x}$K$_x$)Fe$_2$As$_2$ (K-Ba122) can only be explained if the gap is very small on some portion of the Fermi surface, ARPES studies on the same material find a large gap amplitude everywhere \cite{ARPESchinese,Ding}, in the experimentally accessible portions of the Fermi surface close to $k_z$=0.
Some of the apparent contradictions may come from the fact that most studies so far have been carried out in a region of the phase diagram close to the antiferromagnetic (AFM) phase, introducing an additional level of complexity.
One of the few iron-arsenide systems which at present allows a systematic study over the whole doping range, including the strongly overdoped regime, is Ba(Fe$_{1-x}$Co$_x$)$_2$As$_2$ (Co-Ba122) \cite{Athena,NiNiCo}.

Here we report measurements of the thermal conductivity $\kappa$ of Co-Ba122 throughout the superconducting region.
By going down to 50 mK we are able to detect any zero-energy quasiparticles that would exist if there are nodes in the gap.
We find a negligible residual linear term $\kappa_0/T$ in $\kappa/T$ as $T \to 0$ for all doping levels studied, which is strong evidence that there are no nodes in the superconducting gap of Co-Ba122, for ${\bf k}$ directions in the $ab$-plane.
However, the ease with which a magnetic field $H$ applied along the $c$-axis induces a finite $\kappa_0/T$ varies dramatically across the phase diagram. Our results show that the superconducting gap evolves from being uniformly large everywhere on the Fermi surface at low doping to having a very small value somewhere on the Fermi surface at high doping.
This reveals a nodeless yet highly $k$-dependent in-plane gap outside the AFM/orthorhombic phase.


\begin{figure} [t]
\centering
\includegraphics[width=8.5cm]{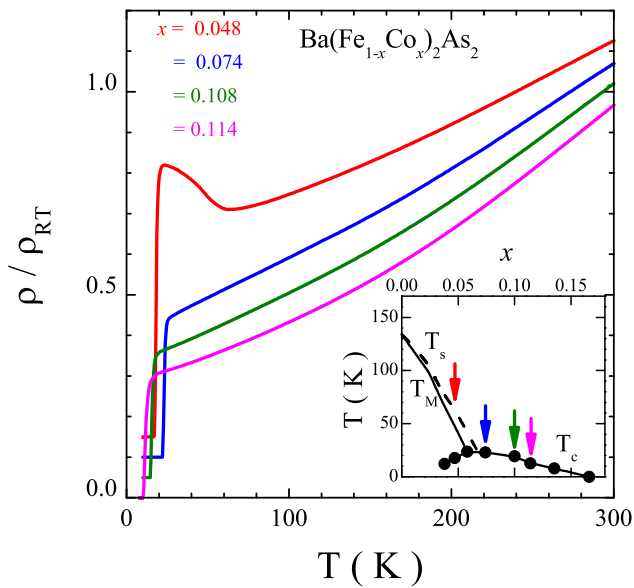}
\caption{\label{resistivity}
Temperature dependence of the in-plane resistivity $\rho(T)$ of the four single crystals of Ba(Fe$_{1-x}$Co$_x$)$_2$As$_2$ studied in this work,
with Co concentrations $x$ as indicated.
The curves are normalized to their room-temperature value $\rho_{\rm RT}$ and rigidly offset by 0.05 for clarity.
Inset: phase diagram showing the orthorhombic phase below $T_S$, the antiferromagnetic phase below $T_{\rm M}$ and the superconducting phase below $T_c$ (from \cite{NiNiCo,McQueeny}),
with arrows indicating the composition of the four samples.
}
\end{figure}



Single crystals of Co-Ba122 were grown from FeAs:CoAs flux, as described elsewhere \cite{NiNiCo}. The doping level in the crystals was determined by wavelength dispersive electron probe microanalysis, which gave a Co concentration $x$ roughly 0.7 times the flux load composition (or nominal content).
We studied four compositions: $x$~=~0.048 (underdoped), 0.074 (optimally doped), 0.108 (overdoped), and 0.114 (overdoped).
The superconducting transition temperature (where $\rho=0$) for each sample is $T_c$~=~17.1, 22.2, 14.6, and 10.1~K, respectively.
Samples were cleaved into rectangular bars with typical size 2 - 3 $\times$ 0.3 $\times$ 0.05~mm$^3$.
Silver wires were attached to the samples with a silver-based alloy,
producing contacts with ultra-low resistance ($< 100~\mu\Omega$).
Thermal conductivity was measured along the [100] direction in the orthorhombic/tetragonal basal ($ab$) plane in a standard
``one-heater-two-thermometer'' technique.
The magnetic field $H$ was applied along the [001] direction ($c$-axis).
All measurements were done on warming after cooling in constant $H$ from above $T_c$ to ensure a homogeneous field distribution.


\begin{figure} [t]
\centering
\includegraphics[width=8.5cm]{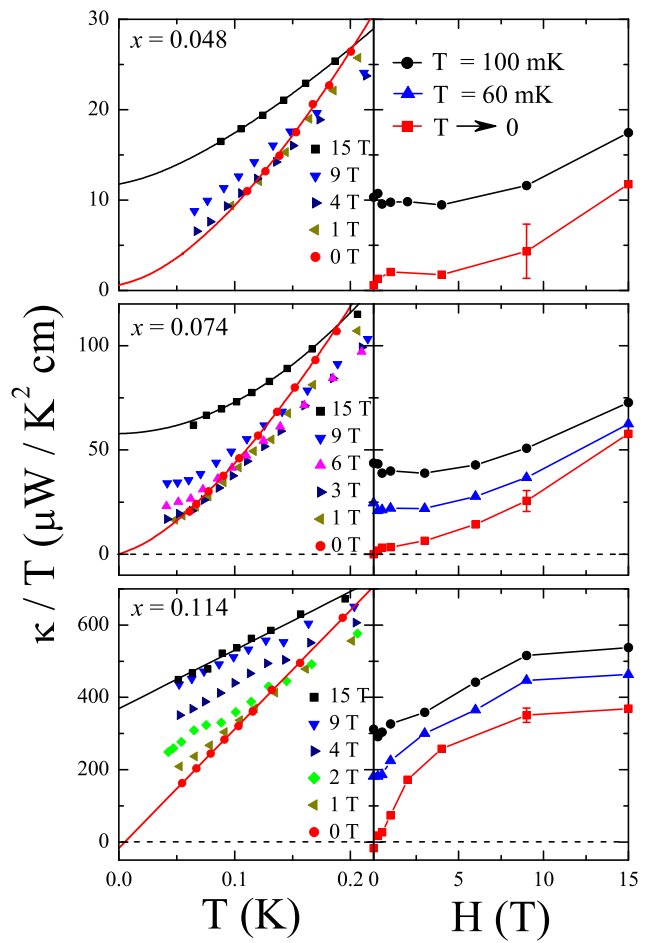}
\caption{\label{6panel}
Left panels:
Temperature dependence of the thermal conductivity, plotted as $\kappa /T$ vs $T$, for three Co concentrations,
each measured in magnetic fields as indicated.
The lines are a power law fit of the form $\kappa/T = a + b T^{\alpha}$ to the $H$~=~0 and $H$~=~15~T data, used to extract the residual linear term $\kappa_0/T \equiv a$ in the $T \to 0$ limit (see text).
Right panels:
field dependence of $\kappa/T$ plotted for three temperatures, as indicated. The $T \to 0$ data (red squares) are obtained from the power-law extrapolations,
with a typical error bar as shown.
}
\end{figure}



In Fig.~\ref{resistivity}, we show the temperature-dependent electrical resistivity $\rho(T)$ of our samples.
In the $x$~=~0.048 sample, the upturn below 60~K signals the onset of tetragonal-to-orthorhombic structural transition at $T_{\rm S}$ followed by the AFM order at a temperature $T_{\rm M}$ \cite{McQueeny}.
No such feature is seen in the other samples.
A smooth extrapolation of $\rho(T)$ to $T = 0$ yields the residual resistivity $\rho _0$,
equal to 203, 62, 59 and 59 $\mu \Omega$~cm for $x = 0.048$, 0.074, 0.108 and 0.114, respectively.
$\rho _0$ is used to determine the normal-state thermal conductivity $\kappa_{\rm N}/T$ in the $T \to 0$ limit via the Wiedemann-Franz law, $\kappa_{\rm N}/T = L_0 /\rho _0$, where $L_0$= 2.45 $\times$~10$^{-8}$ W~$\Omega$~/~K$^2$.
Because the same contacts are used for electrical and thermal measurements, there is no relative uncertainty between the measured $\kappa$ and this electrically-determined
$\kappa_{\rm N}$, other than the small uncertainty in extrapolating $\rho(T)$ to $T = 0$.
%



The thermal conductivity $\kappa(T)$ of the underdoped, optimally doped and most overdoped samples is shown in the left panels of Fig.~\ref{6panel}, for
a magnetic field ranging from $H=0$ to 15~T.
The data below 0.25 K is well described by the function $\kappa/T = a + b T^{\alpha}$, with $\alpha \approx 1 - 1.5$.
The first term, $a \equiv \kappa_0/T$, is the electronic residual linear term of interest here \cite{NJP2009}.
The second term is due to phonons, scattered mostly by crystal boundaries at those low temperatures \cite{Li2008}.
We attribute the fact that this second term is much smaller in the $x = 0.048$ sample, even though all samples have comparable size, to the presence of structural domains produced by the tetragonal to orthorhombic transition at this doping \cite{domains}. The size of domains decreases to $\sim$ 1~$\mu m$ close to a compositional boundary $x_{\rm S}$=0.073$\pm 0.03$ (at which $T_{\rm S}(x) \to 0$) \cite{domains2}, and diminishes effective phonon mean free path. The optimally doped sample $x$=0.074 is located at the border of the orthorhombic phase and its phonon conductivity is still notably suppressed.

The magnitude of the residual linear terms extracted from the fits to the $H=0$ data in the left panels of Fig.~\ref{6panel} are extremely small, below 5 $\mu$W / K$^2$~cm in all cases.
This is within the absolute error bar of $\pm~5$ $\mu$W / K$^2$~cm on $\kappa_0/T$ obtained in similar measurements on samples where no residual linear term is expected \cite{Li2008}.
Within those error bars, all Co-Ba122 samples exhibit a negligible residual linear term, irrespective of the doping level.
Similarly negligible values of $\kappa_0/T$ were observed in K-Ba122 \cite{BaKthermcond} and, as we argued there, this allows us to
safely conclude that the superconducting gap of Co-Ba122 does not contain a line of nodes anywhere on the Fermi surface, at any doping.
By comparison, in a $d$-wave superconductor like the cuprate Tl$_2$Ba$_2$CuO$_{6 + \delta}$ (Tl-2201) with $T_c \simeq 15$~K, $\kappa_0/T = 35$~\% of $\kappa_N/T$ \cite{Proust2002}, while in all four Co-Ba122 samples, $\kappa_0/T < 1 - 2$~\% of $\kappa_N/T$.
This rules out $d$-wave symmetry even for the overdoped samples of Co-Ba122 and further confirms that cuprates and iron arsenides have distinct superconducting states.

The possibility of point nodes in the gap of Co-Ba122 was raised in the context of penetration depth measurements which gave $\Delta \lambda \propto T^{\beta}$ with $\beta \approx$~2 \cite{Gordon}. Zero-energy quasiparticles associated with symmetry-imposed point nodes give a residual linear term in $\kappa$ which grows with impurity scattering (if the gap grows linearly from the node) and for high scattering rates $\kappa_{0}/T$ should become a substantial fraction of $\kappa_{\rm N}/T$ \cite{Graf1996}.
(If the gap grows quadratically from the node, $\kappa_0/T$ is independent of scattering rate and comparable to the case of a line node \cite{Graf1996}.)
The normal-state impurity scattering rate $\Gamma_0$ of Co-Ba122 can be estimated from $\rho_0$ and
the plasma frequency $\omega_{\rm p} = c / \lambda_0$ via $\Gamma_0 = (\omega_{\rm p}^2 / 8 \pi) \rho_0$,
where $\lambda_0$ is the penetration depth at $T \to 0$ and $c$ is the speed of light.
At optimal doping, $\lambda_0 \simeq 200$~nm \cite{Gordon}, so that $\hbar \Gamma_0 / k_{\rm B} T_c \simeq  1 - 2$ in the samples $x = 0.074$, 0.108, and 0.114.
This is very substantial, and would give a large $\kappa_0/T$ from point nodes, comparable to the case of a line node.
We therefore conclude that symmetry-imposed point nodes in the gap of Co-Ba122 are also ruled out, unless they are located in the $k_z$ direction (along the $c$-axis).

In a recent study, a large $\kappa_0/T$ was reported \cite{Machida2009} in a Co-Ba122 sample with $T_c$ and residual resistivity ratio similar to our $x = 0.074$ sample, amounting to 60~\% of $\kappa_{\rm N}/T$, compared with less than 1-2~\% in our sample.
This stands in contradiction not only with our results on more than 10 crystals of Co-Ba122 at various dopings, but also with previous results on
K-Ba122 \cite{BaKthermcond} and Ni-doped BaFe$_2$As$_2$ (Ni-Ba122) \cite{Shiyan}, where in all cases $\kappa_0/T \simeq 0$ at $H=0$.



\begin{figure}
\centering
\includegraphics[width=8.5cm]{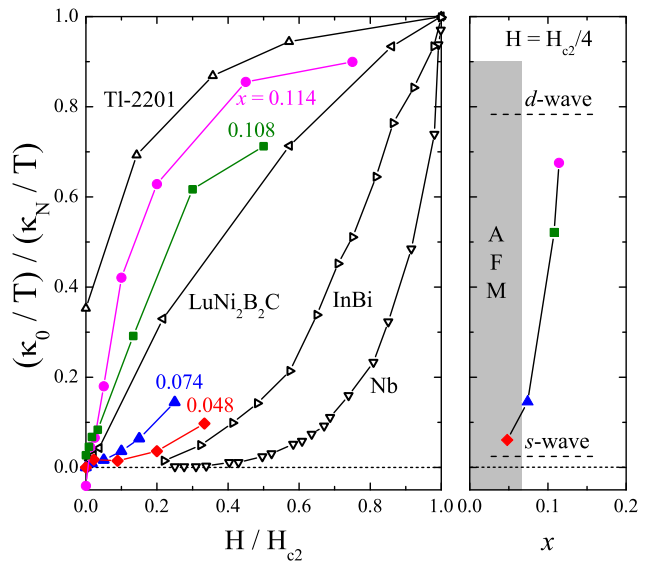}
\caption{\label{comparison}
Left panel:
Residual linear term $\kappa_0/T$ in the thermal conductivity of Co-Ba122 (full symbols) as a function of magnetic field $H$, plotted on normalized scales as $(\kappa_0/T)/(\kappa_{\rm N}/T)$ vs $H/H_{c2}$, where $\kappa_{\rm N}$ is the normal-state thermal conductivity in the $T \to 0$ limit,
obtained from the Wiedemann-Franz law (see text),
and $H_{c2}$ is the upper critical field, for the four Co concentrations $x$ as indicated.
For comparison, we also show the behaviour of pure (Nb) and dirty (InBi) $s$-wave superconductors (from \cite{Li2007}), and of the $d$-wave superconductor Tl-2201
($T_c = 15$~K and $H_{c2} \simeq 7$~T) (from \cite{Proust2002}).
The data for the borocarbide superconductor LuNi$_2$B$_2$C ($T_c = 16$~K and $H_{c2} = 7$~T) is also shown (from \cite{Boaknin2001}).
Right panel:
$(\kappa_0/T)/(\kappa_{\rm N}/T)$ at $H/H_{c2} = 0.25$ in Co-Ba122 as a function of $x$.
This measures the extent to which zero-energy quasiparticles are excited by a relatively small magnetic field compared to $H_{c2}$.
The Co-Ba122 data shows a degree of excitation typical of $s$-wave superconductors in the region of coexisting antiferromagnetic/orthorhombic order at low doping,
rising rapidly to a level typical of a $d$-wave superconductor at higher doping.
}

\end{figure}



The dependence of $\kappa$ on magnetic field $H \parallel c$ is displayed in the right panels of Fig.~\ref{6panel}.
The purely electronic component is given by the extrapolations to $T = 0$ ($T \to 0$; red squares).
To confirm that the $H$ dependence is not affected by the extrapolation procedure, we compare the extrapolated data with cuts through the actual data at two fixed temperatures, namely 60 mK and 100 mK. We see that the latter cuts produce what is essentially a rigid shift of the data relative to
$T \to 0$, caused by the added phonon contribution.
With this validation of the $T \to 0$ data, we plot it on normalized scales in Fig.~\ref{comparison} as $(\kappa_0/T)/(\kappa_{\rm N}/T)$ vs $H/H_{c2}$,
where $H_{c2}$ is the upper critical field, obtained from resistivity measurements in high fields \cite{NiNiCo,Kano2009}, which give
$H_{c2} \simeq 45$, 60, 30, and 20~T for $x = 0.048$, 0.074, 0.108, and 0.114, respectively.
%
This plot reveals the ease with which zero-energy quasiparticles are excited by a magnetic field.
At $x = 0.048$, Co-Ba122 shows the slow rise with upward curvature typical of isotropic $s$-wave superconductors like Nb or InBi (see Fig.~\ref{comparison}), where
the rise in $\kappa_0/T$ is exponentially slow because it relies on the tunneling of quasiparticles between localized states inside adjacent vortex cores,
which at low fields are far apart.
By contrast, at $x = 0.114$, Co-Ba122 shows a very rapid rise with downward curvature more typical of $d$-wave superconductors such as Tl-2201 (see Fig.~\ref{comparison}).
Even though the superconducting gap does not go to zero anywhere, as established by the fact that $\kappa_0/T \simeq 0$, it must become very small somewhere on the Fermi surface to account for such a rapid rise with field.
We can estimate roughly the minimum value of the gap $\Delta$ from the fact that at $x = 0.114$ the rise in $\kappa_0/T$ from zero starts around $H \simeq 0.5$~T
(see lower right panel in Fig.~\ref{6panel}), {\it i.e.} at $H/H_{c2} \simeq 1/40$. Given that $H_{c2} \propto \Delta^2$, we obtain for the gap minimum
$\Delta_{\rm min} \simeq \Delta_0/6$,
in terms of the gap maximum, $\Delta_0$, which controls $H_{c2}$.

This is similar to what has been found in the borocarbide superconductor LuNi$_2$B$_2$C ($T_c = 16$~K and $H_{c2} = 7$~T), where even though
$\kappa_0/T = 0$ at $H = 0$, $\kappa_0/T$ rises very rapidly at low fields \cite{Boaknin2001} (see Fig.~\ref{comparison}).
There, an estimate of the gap minimum gave $\Delta_{\rm min} \simeq \Delta_0/10$ \cite{Boaknin2001}.
Recent ARPES studies on the closely related material YNi$_2$B$_2$C found a sharp superconducting gap minimum at positions on the Fermi surface which are connected by a nesting wavevector \cite{ARPES-YNi2B2C}.

The fact that the deep minimum in the gap of Co-Ba122 is only present in composition free of competing orders (in our case orthorhombic/AFM phase) is consistent with previous heat transport studies
performed at single dopings on other arsenide superconductors.
There is no AFM order in K-Ba122 at $x \simeq 0.3$, where a relatively rapid increase of $\kappa_0/T$ vs $H$ was reported \cite{BaKthermcond}, indicative of a gap minimum.
In the low-$T_c$ nickel-arsenide superconductor BaNi$_2$As$_2$ ($T_c = 0.7$~K) \cite{Kurita}, which orders around 132 K with an as yet unknown order parameter \cite{Ronning}, the weak field dependence of $\kappa$ is consistent with a full gap everywhere on the Fermi surface, compatible with the behavior in underdoped Co-Ba122.
In Ba(Fe$_{1-x}$Ni$_x$)$_2$As$_2$ (Ni-Ba122), only a weak $H$ dependence was observed near optimal doping ($T_c = 20$~K) \cite{Shiyan} --
it will be interesting to see how that evolves at higher doping. Note that in Ni-Ba122, the competing magnetic/structural phase may extend to higher electron doping \cite{Canfield-doping}.
In the low-$T_c$ iron-arsenide superconductor LaFePO ($T_c = 7$~K), a line of nodes in the gap has been inferred from the linear $T$ dependence of the penetration depth \cite{Fletcher}. Such a line of nodes would necessarily lead to a finite $\kappa_0/T$, and indeed it does, as reported recently \cite{Matsuda}.
This extreme form of anisotropy again occurs in the absence of competing orders.


In summary, zero-field heat transport in the $T \to 0$ limit shows that the superconducting gap of Co-Ba122 does not go to zero at any point on the Fermi surface, for quasiparticle $k$ vectors in the basal plane. This is true throughout the phase diagram, from underdoped to overdoped, and is
consistent with heat transport data in both K-Ba122 \cite{BaKthermcond} and Ni-Ba122 \cite{Shiyan}, measured previously at a single doping close to optimal.
In Co-Ba122, the initial rise of $\kappa_0/T$ with magnetic field $H$ varies enormously with doping, from slow and typical of an isotropic $s$-wave gap at low doping,
in the region of coexisting AFM/orthorhombic order, to very fast and comparable to a $d$-wave gap at high doping.
This fast initial rise shows that the gap must be very small on some portion of the Fermi surface, roughly a sixth of the gap maximum.
One way to understand this pronounced $k$-dependence of a nodeless superconducting gap is in terms of an extended $s$-wave scenario \cite{Wang2009,Mishra2009,Chubukov2009},
whereby the symmetry of the order parameter is $s$-wave but the pairing interaction is so strongly $k$-dependent that the gap is highly anisotropic, going to zero in certain directions.
(In addition, it may also change sign, as in the $s_{\pm}$ state \cite{Mazin2008}.)
Because these gap zeros or nodes are `accidental', {\it i.e.} not imposed by symmetry (as they would be for a $d$-wave order parameter),
the effect of impurity scattering is not to broaden them (as in a $d$-wave gap), but to lift them, making the gap more isotropic \cite{Mishra2009}.
This could then explain why the undoped material LaFePO has a line of nodes, while the
doped superconductor Co-Ba122 doesn't have nodes, given that the former has a 10 times smaller impurity scattering rate (as measured by the residual resistivity ratio).
In other words, as recently proposed \cite{Mishra2009}, the pronounced $k$-dependence of the gap produces nodes in the pure limit which turn into deep minima in the presence of strong impurity/disorder scattering.
The fact that no pronounced anisotropy ({\it i.e.} field dependence of $\kappa$) is observed in Co-Ba122 when there is AFM order leads us to propose that the AFM order in iron arsenides reconstructs the Fermi surface by gapping those regions where the minima in the superconducting gap are located.
This is reminiscent of what occurs in the borocarbide superconductors $R$Ni$_2$B$_2$C ($R=$~Y or Lu) \cite{Boaknin2001}, where gap minima are observed at $k$-locations connected by a nesting wavevector \cite{ARPES-YNi2B2C}.
This suggests an intimate interplay of magnetism and superconductivity.


Work at the Ames Laboratory was supported by the Department of Energy-Basic Energy Sciences under Contract No. DE-AC02-07CH11358.
R.P. acknowledges support from the Alfred P. Sloan Foundation.
L.T. acknowledges support from CIFAR, NSERC, CFI and FQRNT.


\end{document}